\documentclass[letter]{ptptex}
\usepackage{latexsym}
\usepackage{amsfonts}
\usepackage{amsmath}
\usepackage{amssymb}
\usepackage{graphicx}
\usepackage{wrapft}
\notypesetlogo
\title{The Fate of a Five-Dimensional Rotating Black Hole \\
via Hawking Radiation}

\author{Hidefumi
\textsc{Nomura},$^{1,}$\footnote{E-mail: nomura@gravity.phys.waseda.ac.jp}
\ Shijun \textsc{Yoshida},$^{1,}$\footnote
 {E-mail: shijun@waseda.jp} \
 Makoto \textsc{Tanabe},$^{1,}$\footnote
 {E-mail: tanabe@gravity.phys.waseda.ac.jp} and \\
Kei-ichi  \textsc{Maeda}$^{1,2,3,}$\footnote
{E-mail: maeda@gravity.phys.waseda.ac.jp}%
}

\inst{
$^{1}$Department of Physics, Waseda University,
Tokyo 169-8555, Japan\\
$^{2}$Advanced Research Institute for Science and Engineering,
Waseda  University,  Tokyo 169-8555, Japan\\
$^{3}$ Waseda Institute for Astrophysics,
Waseda  University,  Tokyo 169-8555, Japan\\
}

\recdate{\today}

\abst{
We study the evolution of a five-dimensional rotating black hole
emitting scalar field radiation via 
the Hawking process
for arbitrary initial values of the two rotation parameters $a$ and $b$.  
It is found that any such black hole whose initial rotation 
parameters are both nonzero  evolves toward an asymptotic state 
 $a/M^{1/2}=b/M^{1/2}={\rm const}(\neq 0)$, where this constant
 is independent of the initial values of $a$ and $b$.
}

\begin{document}

\maketitle

The conventional view of black hole evaporation is that, regardless
of its initial state, Hawking radiation will cause a black hole to
approach an uncharged, zero angular momentum state long before
all its mass has been lost. For this reason, in some works, 
it is assumed that  as a black hole 
evaporates close to the Planck scale, where quantum gravity is 
required to determine its  evolution, the final asymptotic state 
is  described by Schwarzschild solution.	

However, Chambers, Hiscock and Taylor \cite{Chambers} investigated, in some 
detail, the evolution of a Kerr black hole emitting 
scalar field radiation via
the Hawking process, and showed that the ratio  of the 
black hole's specific angular momentum to its mass,  $\tilde{a}=a/M$, evolves
toward a stable nonzero value ($\tilde{a}\rightarrow 0.555$).
This means that a rotating black hole will evolve toward a final state
with non-zero angular momentum if there is a scalar field.
In this Letter, we shall extend the analysis of Chambers, Hiscock and Taylor
to a higher-dimensional case for the reasons  described below.  Considering 
the five-dimensional case specifically, we investigate the evolution of 
a five-dimensional rotating Myers-Perry (MP) black hole \cite{myers}
with two rotation parameters through scalar field radiation.

Recently,  black holes in N ($\ge 4$) dimensions have  
attracted much attention. 
This is due to interest in the brane world scenario \cite{arkani,randall}.
From a phenomenological point of view, the most exciting possibility
for the brane world scenario  is
that it might be possible to produce  
higher-dimensional mini-black holes in particle colliders, 
such as the CERN Large Hadron Collider (LHC), 
or to find them in cosmic ray events \cite{Anchordoqui}. 
A  black hole  produced in this manner would evaporate rapidly and emit many 
standard particles. 
Hence it would lese most
of its mass and angular momenta through  Hawking radiation
\cite{Hawking-creation} or superradiance, which is intrinsic
to a rotating or charged black hole \cite{Unruh,Gibbons}.
A few hot quanta
emitted in the final Planck phase, which cannot be treated
semiclassically, would not consist of the main part of the decay
products \cite{Giddings}.
In most of the literature,
the ``spin-down" phase of black hole evolution, in which a black hole loses
its angular momenta, is simply ignored, and a Schwarzschild black
hole is assumed.

In generic particle collisions, however, the impact parameter will be
non-zero. Therefore, most black holes produced in a collider would be rotating
and could be described by a higher-dimensional MP solution \cite{myers}, or 
other rotating objects, such as a black ring \cite{emparan}.
For this reason, we  focus on a ``spin-down" phase through scalar field radiation.
A five-dimensional rotating black hole possesses three Killing vectors:
$\partial t$, $\partial \phi$, and $\partial \psi$.
Therefore a five-dimensional black hole has two rotation parameters.
For a five-dimensional MP black hole with
one rotation parameter, Ida,  Oda and  Park \cite{Ida}
found the formulae for the black body factor in a low-frequency
expansion and the power spectra of the Hawking radiation.
However, if a brane is not infinitely thin but, rather, has a thickness 
in the order of
a fundamental scale ($\sim$ TeV),
we  expect to exist a second component of angular momentum.
We therefore study the case of two rotation parameters.
Frolov and Stojkovi$\acute{\rm c}$ first derived expressions for the energy and
angular momentum fluxes from a five-dimensional rotating black hole
with two rotation parameters \cite{frolov}.
In this work, we numerically  evaluated the
 quantum radiation  from a
five-dimensional rotating black hole
with two rotation parameters, $a$ and $b$, which we assume to be
positive, without loss of generality.
We found that  such a black hole evolves toward an asymptotic state
characterized by a stable values $a=b\sim 0.1975\,(8 M/3\pi)^{1/2}$, where $M$
is the mass of the black hole. 
We also show that the asymptotic state can be described by 
$a\sim 0.1183\,(8 M/3\pi)^{1/2}$ and $b=0$ if one of the initial rotation parameters is 
exactly zero. 

We start with the quantum radiation of a massless scalar field $\Phi$,
which is minimally coupled, for a five-dimensional
MP black hole with two rotation parameters \cite{frolov},
(see also Ref. \citen{nomura} for  details).
To quantize the scalar field, we expand it as
$
\Phi=R(r)\Theta(\theta) e^{im\phi}
e^{in\psi} e^{-i\omega t}.
$
For the vacuum state, we adopt the (past) Unruh vacuum state
$| U^- \rangle $, which mimics the state of collapse of a 
star to a black hole \cite{Unruh}.
Calculating the vacuum expectation value of the energy-momentum tensor
of the scalar field, we can evaluate the emission rates of
the total energy and angular momenta, which give
the changes of the black hole mass $M$ and angular momenta $J_\phi$ and
$J_\psi$ as 
\begin{eqnarray}
\dot{M}&=&-\pi\sum _{lmn}
\int_{0}^{\infty}d\omega
~\frac{\omega^2}{\omega_{+}}
\frac{\Gamma_{lmn} }
{e^{2\pi\omega_{+}/\kappa}-1} \label{eq:dE/dt}, \\
\dot{J}_\phi&=&-\pi\sum_{lmn}
\int_{0}^{\infty}
d\omega ~
\frac{m\omega}{\omega_{+}}
\frac{\Gamma_{lmn}  }
{e^{2\pi\omega_{+}/\kappa}-1} \label{eq:dJ_phi/dt}, \\
\dot{J}_\psi&=&-\pi\sum_{lmn}
\int_{0}^{\infty}
d\omega ~
\frac{n\omega}{\omega_{+}}
\frac{\Gamma_{lmn}  }
{e^{2\pi\omega_{+}/\kappa}-1} \label{eq:dJ_psi/dt},
\end{eqnarray}
where $\omega_+=\omega-m\Omega_\phi-n\Omega_\psi$, 
$\kappa=(r_+^2-r_-^2)/2Mr_+$,
$l$ is the eigenvalue of the angular function $\Theta(\theta)$, 
and $\Gamma_{lmn}$ is the greybody factor, which
is identical to the absorption probability of the incoming
wave of the corresponding mode.
The  values $r_+$ and $r_-$ represent the event horizon
 and the inner horizon of the black hole,
 respectively.
The quantities $\Omega_\phi=a/(r_+^2+a^2)$ and $\Omega_\psi=b/(r_+^2+b^2)$ 
are the two angular velocities 
at the horizon $r_+$.  The superradiance modes are given by
the condition $0<\omega<m\Omega_\phi+n\Omega_\psi$.
From this condition, we find the  interesting feature that
a counter-rotating particle can be created by  superradiance (i.e.
if $\Omega_\phi\gg\Omega_\psi$ and $m\geq 1$) because
 the superradiance condition is satisfied
for a counter-rotating particle ($n<0$) (see Ref. \citen{nomura} for  details).

Using the above formula for the quantum creation of a scalar field, 
we investigate the evolution of a five-dimensional MP black hole
with two rotation parameters.
From the condition for the existence of  horizon(s), we obtain the condition
$a+b \le r_s$ constraining the angular momenta, 
where $r_s$ is a typical scale length which is related to the 
gravitational mass $M$ of the  black hole as $r_s^2=8M/3\pi$.

As shown by Page \cite{Page}, it is convenient to introduce  scale 
invariant rates of change for the mass and angular momenta of 
an evaporating black hole as 
\begin{eqnarray}
f\equiv-r_s^2\dot{M}, \quad
g_a\equiv-\frac{r_s}{a_*}\dot{J_\phi}, \quad {\rm and} \quad
g_b\equiv-\frac{r_s}{b_*}\dot{J_\psi},\label{emission_rateJpsi}  
\end{eqnarray}
where $a_*=a/r_s$ and $b_*=b/r_s$. In terms of the 
scale invariant functions $f$, $g_a$, and $g_b$, the time evolution  
equations for $a_*$ and $b_*$ are given by 
\begin{eqnarray}
\frac{\dot{a_*}}{a_*}=-\frac{8}{3\pi}\frac{fh_a}{r_s^4} \quad
{\rm and} \quad
\frac{\dot{b_*}}{b_*}=-\frac{8}{3\pi}\frac{fh_b}{r_s^4},  
\label{eq:evolution-eq}
\end{eqnarray}

where the dimensionless functions $h_a$ and $h_b$ are defined as 
\begin{eqnarray}
h_a \equiv \frac{d \ln a_*}{d \ln M}=\frac{3}{2}\left(\frac{g_a}{f}-1
\right) 
\quad
{\rm and} \quad
h_b \equiv \frac{d \ln b_*}{d \ln M}=\frac{3}{2}\left(\frac{g_b}{f}-1
\right).
\end{eqnarray}

We now discuss the evolution of $a_*$ and $b_*$, 
as determined through the numerical evaluation of
 $f$, $g_a$ and $g_b$.
Henceforth, we use units such that $r_s=1$.
In the dynamical system (\ref{eq:evolution-eq}), 
a fixed point plays an important role.
It is defined by $h_a=0$ and $h_b=0$.
Note that $f$ is positive definite.
If $h_a$ ($h_b$) is positive, 
then $a_*$ ($b_*$) decreases, while if $h_a$ ($h_b$)  negative, 
then $a_*$ ($b_*$) increases.
Because $h_a$ ($h_b$) depends not only on  $a_*$ ($b_*$) but also 
on $b_*$ ($a_*$), $h_a=0$ ($h_b=0$) gives a curve in 
the $a_*$-$b_*$ plane.
Since there is symmetry between $a_*$ and $b_*$, the fixed point
should be  symmetrical, too. 

\begin{figure}[htb]
\parbox{\halftext}
{
\includegraphics[height=5cm,clip]{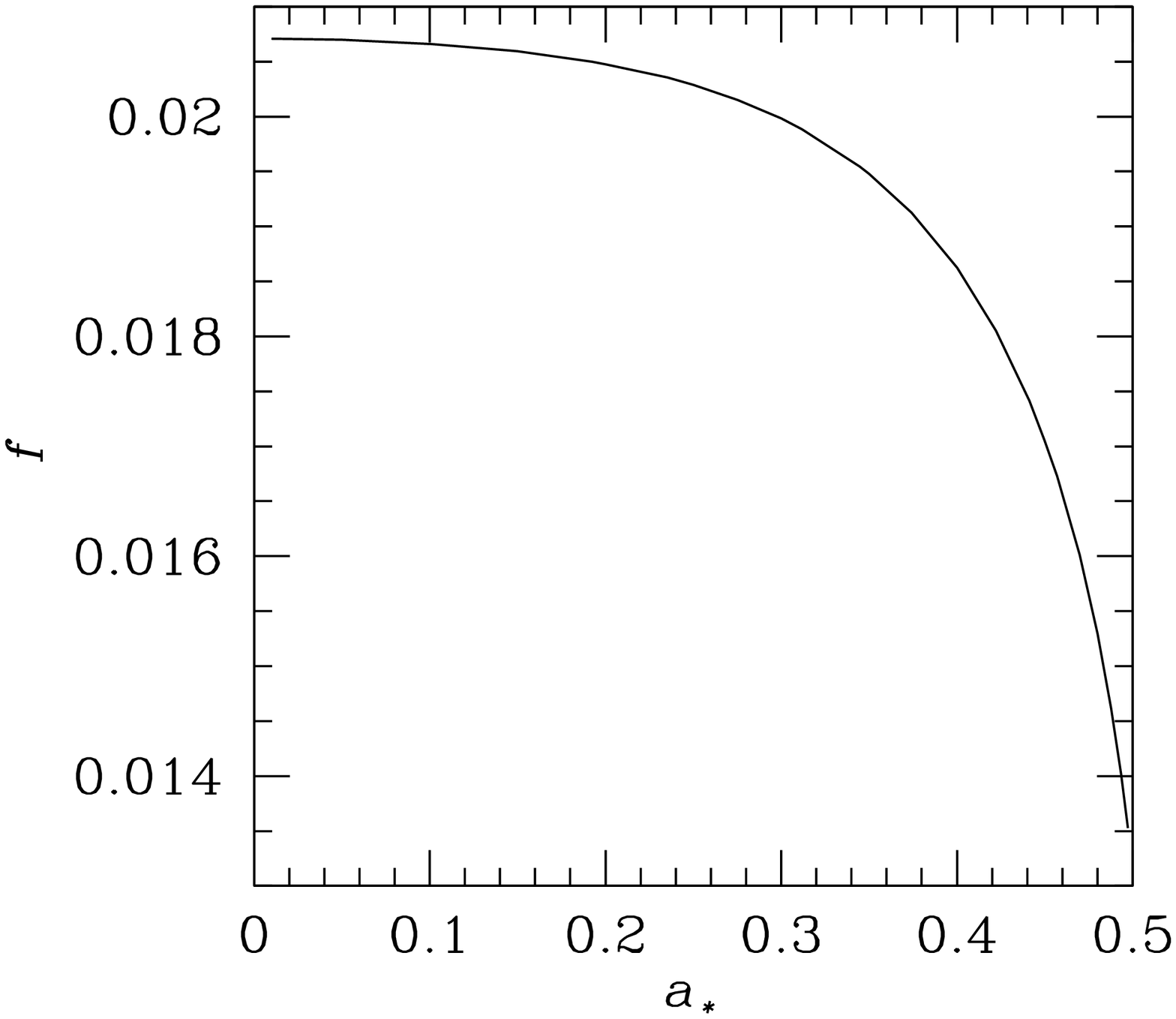}
\caption{The scale invariant quantity $f$, which represents
the mass loss rate, as a function of $a_*$ for
the case  $a_*=b_*$. The function $f$ is positive definite 
by definition.}
\label{f5}}
\hfill
\parbox{\halftext}
{
\includegraphics[height=5cm,clip]{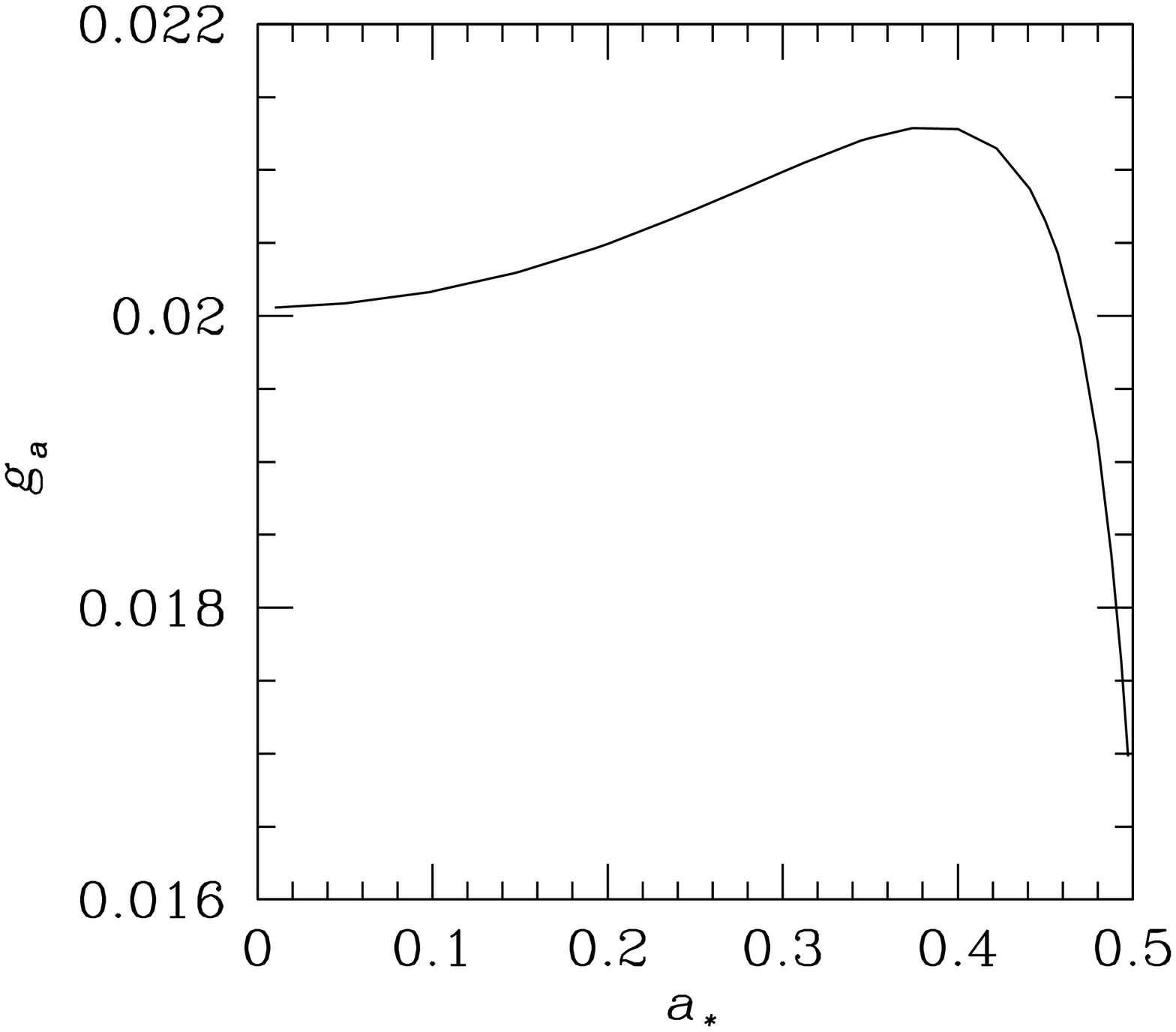}
\caption{The scale invariant quantity $g_a$, which represents 
the loss rate of the angular momentum $J_\phi$, as a function of
  $a_*$ for
the case  $a_*=b_*$, for which $g_a=g_b$. The function $g_a$
is positive definite by definition.}
\label{f6}
}
\end{figure}

\begin{wrapfigure}{l}{6.6cm}
\includegraphics[height=5cm,clip]{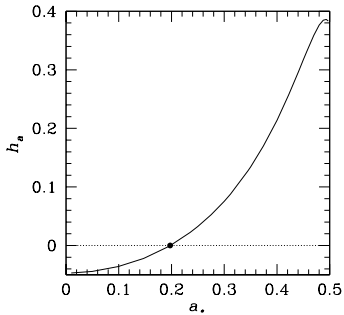}
\caption{The scale invariant quantity $h_a$, which represents 
the change rate of  $a_*$, as a function  of $a_*$ for 
the casef $a_*=b_*$, for which $h_a=h_b$. The function $h_a(a_*)$ 
has a zero at $a_*=a_*^{\rm (cr)}\simeq 0.1975$ (a black spot).}  
\label{f4}
\end{wrapfigure}

We first discuss the
behavior of the mass and angular momentum loss rates 
in the case   $a=b$ (and hence $a_*=b_*)$.
Fig. \ref{f5} displays the mass  loss rate $f(a_*)$ in terms of $a_*$ $(=b_*)$.
The mass loss rate through the scalar radiation is more effective at smaller values of
 $a_*$.  We  depict the angular momentum loss rate $g_a(a_*)$ $(=g_b(a_*))$
in Fig. \ref{f6}.
The function $g_a(a_*)$ has a maximum at $a_*=a_*^{\rm (max)}\approx 0.3844$.
We plot the function $h_a(a_*)$ $(=h_a(a_*))$ in
Fig. \ref{f4}.
We find $h_a(a_*)=0$ at
$a_*=a_*^{\rm (cr)}\approx 0.1975$, which is a fixed point in the present
 dynamical system.
An important property of the function $h_a(a_*)$ is
that $h_a(a_*)<0$ [$h_a(a_*)>0$] for $a_*<a_*^{\rm (cr)}$ [$a_*>a_*^{\rm (cr)}$].

As a result, the fixed point $(a_*,b_*)
=(a_*^{\rm (cr)},a_*^{\rm (cr)})$ is stable along the line $a_*=b_*$.
Hence, a black hole formed with equal rotation parameters, $a_*=b_*\ne 0$, 
 will eventually
 reach an asymptotic state characterized
by $a_*$ $(=b_*)=a_*^{\rm (cr)}$, through  scalar field radiation.

\begin{figure}[htbp]
\parbox{\halftext}{
\includegraphics[height=5cm,clip]{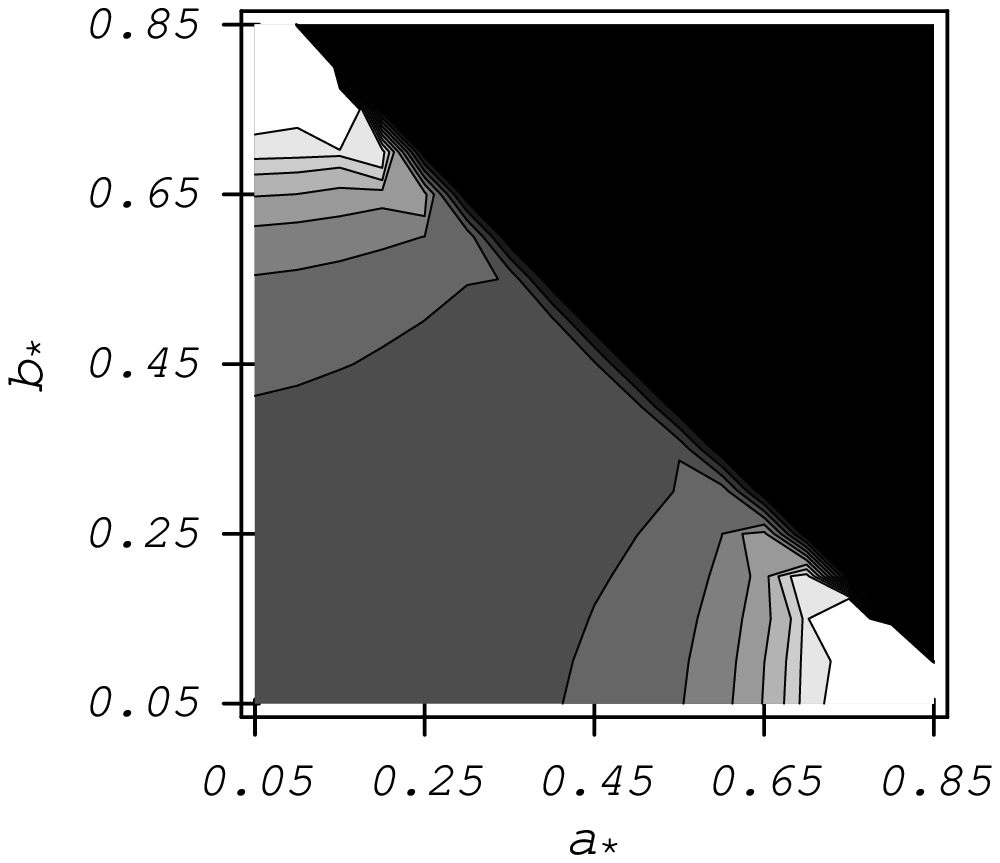}
\caption{The contours of  $f$ in the $a_*$-$b_*$ plane. 
The darkest and  brightest regions correspond to 
zero and $f_{\rm max}$ (the maximum of $f$),
which is given by 
$f_{\rm max}\simeq f(0.85,0.05)=4.349$,
 respectively. The difference between 
two contours is  $f_{\rm max}/10$.
The black region is  forbidden, because there is no horizon in this
region.}
\label{f1}}
\hfill
\parbox{\halftext}{
\includegraphics[height=5cm,clip]{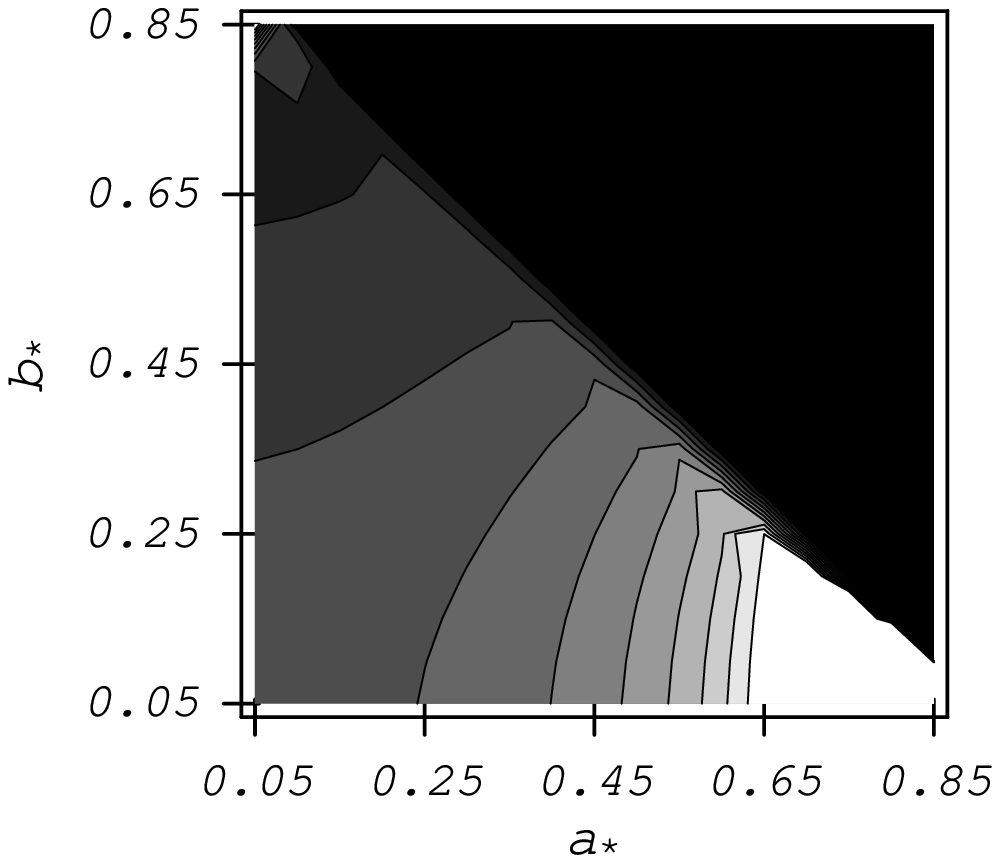}
\caption{The contours of  $g_a$ in the $a_*$-$b_*$ plane. 
The black and  white regions correspond to 
zero and $g_{a,{\rm max}}$  (the maximum of $g_a$),
which is given by 
$g_{a,\rm max}\simeq f(0.85,0.05)=5.92467$,
 respectively. The difference between 
two contours is  $g_{a,{\rm max}}/10$.
The black region is  forbidden, because there is no horizon
in this region.}
\label{f7}}
\end{figure}

In order to investigate the more generic case ($a\neq b$),
we have to analyze Eq. (\ref{eq:evolution-eq}).
For this purpose, we depict the contour plots of
 $f$  and $g_a$  in Figs. \ref{f1} and \ref{f7}, respectively. 
($g_b$ is obtained by exchanging
the axes for $a_*$ and $b_*$ in Fig. \ref{f7}.)  

In the $a_*$-$b_*$ plane, the region in which  $a_*+b_*>1$ is
forbidden, because there is no horizon (the black region
in Figs. \ref{f1} and \ref{f7}).
In Fig. \ref{f1}, there are two bright regions
(one for large $a_*$ and small $b_*$, and 
one for small $a_*$ and large $b_*$), where
$f$ becomes large. This means that the creation rate is high in
these regions.
In Fig. \ref{f7}, there is only one bright region
(for large $a_*$ and small $b_*$).
Therefore, the angular momentum  $J_\phi$ is emitted
effectively only in this region.  This is the
superradiance effect.
For the angular momentum $J_\psi$,
if $b_*$ is large, we find  effective emission.
This means that the superradiance modes give a dominant contribution
to the particle creation.

There is one interesting observation here:
If the two rotation parameters are equal (i.e. $a_*=b_*$),
the emission rates are suppressed even if the black hole is in a
maximally rotating state ($a_*=b_*=0.5$).
In the case  $a=b$, something strange seems to happen, 
and the system  behaves like a ``spherically symmetric" black hole.
In fact, the angular equation for $\Theta(\theta)$ in this case
is exactly the same as
that for the Schwarzschild black hole \cite{frolov}.
This may suppress  the superradiance effect.
This  is consistent with the result given in Ref. \citen{Nozawa},
the efficiency of energy extraction for a MP black hole
is very small in the case  that the  rotation parameters
 are equal.

\begin{wrapfigure}{l}{6.6cm}
\includegraphics[height=5cm,clip]{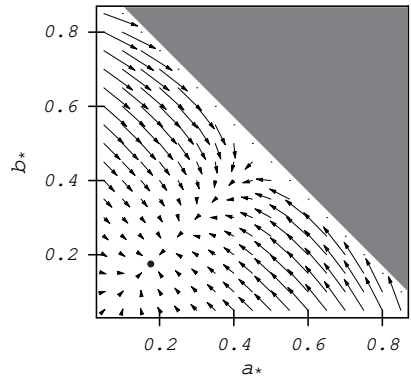}
\caption{
The vector field describes the direction in which $a_*$ and $b_*$
evolve,
i.e. $(\dot{a}_*,\dot{b}_*)$. 
For any initial values of $a_*$ and $b_*$, the system  
evolves toward $a_*=b_*=0.1975$ (the black spot),
which is a stable  fixed point.
The shaded region is  forbidden.}
\label{f11}
\end{wrapfigure}

In order to see the evolution of 
a black hole in the $a_*$-$b_*$ plane,
we plot the vector field $(\dot{a}_*, \dot{b}_*)$ with  arrows
in Fig. \ref{f11}.
From this figure, we see how the values of $a_*$ and $b_*$ 
 evolve toward the fixed point.
We can also prove that the fixed point is a stable attractor
(see Ref. \citen{nomura} for details).

In Fig. \ref{f11} the arrows far from the symmetry line of $a_*=b_*$
are very large. Then, if the initial value of $a_*$ ($b_*$) 
is large, while 
that of $b_*$ ($a_*$) is small, 
 $a_*$ and $b_*$ first approach
the same value.
Near the fixed point  $(a_*^{\rm (cr)}, a_*^{\rm (cr)})$, 
the arrows are very small, which means that 
the evolution toward the fixed point is  slow. 
We thus find that after reaching a state with 
$a_*=b_*$, $a_*$ and $b_*$  eventually 
evolve together toward the fixed point  
 $(a_*^{\rm (cr)}, a_*^{\rm (cr)})\approx (0.1975,0.1975)$.
This means that any rotating black hole with two non-zero rotation 
parameters will evolve toward
a final state with
the same specific angular momenta, $a_*=b_*=a_*^{\rm (cr)}\approx 0.1975$.
For a black hole with only one non-trivial rotation parameter,
i.e. $a \neq 0$ and $b=0$ exactly, 
we obtain the stable fixed point from the equation $h_a(a_*, 0)=0$,
which yields $a_* \approx 0.1183$.

Finally, consider the evaporation time of the black hole.
In the above analysis, we showed that our dynamical system 
(\ref{eq:evolution-eq})
has one stable attractor, which can be 
reached through quantum particle production.
However, the black hole may evaporate away before 
this fixed point is reached.
Whether this happens depends on the evaporation time and the evolution time in the
$a_*$-$b_*$ plane. 
We can evaluate the evaporation time scale $\tau_M$
 using the emission rate of the black hole mass as $
\tau_M =-M/\dot{M}$,
 and we can evaluate the evolution time scale $\tau_{a_*}$
using  the evolution equation (\ref{eq:evolution-eq})
as
$
\tau_{a_*}=a_*/|\dot{a}_*|.
$

We thus find  that $\tau_M/\tau_{a_*}=8|h_a|/(3\pi)\sim O(1)$.
However, this does not mean that the black hole will evaporate away
before reaching the fixed point.
If the integrated evaporation time, which depends on the initial mass 
of the black hole, is much longer than the evolution time,
we have enough time to realize the final state described by 
the fixed point.
Therefore, we conclude that if a black hole has a
 mass that is  
larger than the fundamental Planck mass scale,
its two specific angular momenta will eventually become equal
when it  evaporates away.

%
%
\section*{Acknowledgements}
We would like to thank J. Koga and T. Torii for useful discussions. 
This work was partially supported by the Grant-in-Aid for Scientific Research
Fund of the MEXT (No. 17540268), by the Waseda
University Grant for Special Research Projects, and by 
a Grant for The 21st Century
COE Program (Holistic Research and Education Center for Physics
Self-organization Systems) at Waseda University.

%

\end{document}